\documentclass[revtex,pra,preprint,showpacs]{revtex4}
 \usepackage{bm}

\usepackage{graphicx}
\usepackage{amsmath}
\usepackage{amssymb}

{ \end{list} }

{ \end{list} }

\usepackage{amsmath,amssymb}
\usepackage{dcolumn}
\newcolumntype{.}{D{.}{.}{-1}}

\usepackage{footnote}

\usepackage{longtable}

\usepackage[utf8]{inputenc}
\usepackage[T1]{fontenc} 

\usepackage{comment}
\begin{document}

\title{
Testing excited-state energy density functional and potential with the ionization potential theorem
}
\author{M.~Hemanadhan, Md.~Shamim and Manoj~K.~Harbola}
\address{Department of Physics, Indian Institute of Technology, Kanpur 208 016, India}

\begin{abstract}
The modified local spin density functional and the related local potential for excited states is 
tested by employing the ionization potential theorem. The functional is constructed by splitting $k$-space. 
Since its functional derivative cannot be obtained easily, 
the corresponding potential is given by analogy to its ground-state counterpart.
Further to calculate the highest occupied orbital energy $\epsilon_{max}$ accurately, the potential is corrected for
its asymptotic behavior by employing the van Leeuwen and Baerends correction to it. $\epsilon_{max}$ so obtained is 
then compared with the $\Delta$SCF ionization energy calculated using the MLSD functional. It is shown that the two 
match quite accurately.
\end{abstract}
\maketitle

\section{Introduction}
\label{sec:introd}
Ground-state density functional theory (gDFT) is the most widely used theory for electronic structure calculations
~\cite{book:Parr-Yang:1989,book:Dreizler-Gross:1990,book:March:1992,book:Engel-Dreizler:2011}. 
The key to its success has been accurate exchange-correlation functionals $E_{xc}$ developed over the past few 
decades~\cite{becke:1988,perdew-burke-ernzerhof:1996,Tau-Perdew-etal:2003}.
The exchange-correlation potential $v_{xc}$ required for the self-consistent calculations (SCF) is then obtained either 
by taking functional derivative of the $E_{xc}$ or in some cases by using model 
potentials~\cite{leeuwen-baerends:1994,Umezawa:2006,becke-johnson:2006}.

It is then natural to ask if the ground-state theory can be extended to study excited-states to 
perform self-consistent Kohn-Sham calculations for the density and total energy of excited-states.
Although time-dependent density functional theory  (TDDFT) is now routinely used for calculations 
of excitation energies and the corresponding oscillator strengths, the theory has its 
limitations~\cite{book:Ullrich:2012}. 
On the other hand, the progress of  time-independent excited-state DFT (eDFT) has been slow. 
Some of the earlier work includes the extension of ground-state theory to the lowest energy states of a given symmetry by 
Gunnarsson and Lundqvist~\cite{gunnarsson-lundqvist:1976,gunnarsson-lundqvist:1976:err:1977}, 
Ziegler et al.~\cite{ziegler-rauk-baerends:1977} and von Barth~\cite{barth:1979}.  
Subsequent work are the development of ensemble theory to excited-states by Theophilou~\cite{theophilou:1979}, 
Gross, Oliveira, Kohn~\cite{gross-oliveira-kohn:1988,oliveira-gross-kohn:1988}, 
and its application to study transition energies of atoms by Nagy~\cite{nagy:1996}. 
Recently, the work by G{\"o}rling~\cite{gorling:1999} and Levy and Nagy~\cite{levy-nagy:1999,nagy-levy:2001},
both based on constrained-search approach~\cite{Levy:1979}, rekindled interest in eDFT.
Following this, Samal and Harbola explored density-functional theory for excited-states further
~\cite{harbola:2002,harbola:2004,samal-harbola:2005,samal-harbola:2006,samal-harbola:2006b}.

A crucial requirement for implementing eDFT is the appropriate functionals for the excited-states. 
These functionals should be as easy to use as the ground-state functionals and be such that 
improved functionals can be built upon them.  
For the ground-states such a functional is  provided by the local-density approximation (LDA), 
which is based on the homogeneous electron gas (HEG). 
Motivated by this, we have proposed an LDA-like functional for excited-states~\cite{samal-harbola:2005}. 
This functional is also obtained using the homogeneous electron gas. 
The spin-generalization of the functional, the modified local spin-density (MLSD) functional has been 
shown to lead to accurate transition-energies~\cite{samal-harbola:2005}. 
Encouraged by this, we have been subjecting our method of constructing the functional to more and more 
severe tests~\cite{hemanadhan-harbola:2010,hemanadhan-harbola:2012,shamim-harbola:2010}. 
With this in mind, we test our method for the satisfaction of the ionization potential (IP) theorem in this paper. 

According to the ionization potential (IP) theorem for the 
ground-states~\cite{PhysRevLett.49.1691,Levy-Perdew-Sahni:1984,Katriel-Davidson:1980} or an excited-states,   
the highest occupied Kohn-Sham orbital energy ($\epsilon_{max}$) for a system is 
equal to the negative of the ionization potential $I$~\cite{PhysRevLett.49.1691}. 
Thus 
\begin{equation}
 \epsilon_{max} = -I(N) \equiv    E(N) - E(N-1)
 \label{eq:ip}
\end{equation}
where $E(N)$, and $E(N-1)$ are the energies corresponding to $N$ and $N-1$ electron systems such that 
$I(N)$ is smallest. 
The difference of these energies for the $N$ and $N-1$ electron system calculated self-consistently is referred to as $\Delta$SCF. 
The relationship of Eq.~\eqref{eq:ip} arises because the asymptotic decay of the electronic density of a system is related to 
its ionization potential; on
the other hand, for a Kohn-Sham system it is governed by $\epsilon_{max}$, thereby relating the two quantities.
Thus, if the exact functionals were known, the corresponding Kohn-Sham calculation will give $\epsilon_{max}$, $E(N)$ and 
$E(N-1)$ so that Eq.~\eqref{eq:ip} is satisfied. However, this is not the case when approximate functionals are used.
For instance, when ground-state calculations are done using the LDA, the $\Delta$SCF values are accurate, but the 
$\epsilon_{max}$ are roughly $50\%$ of the $\Delta$SCF energy or the experimental values~\cite{w2012crc}.
This is due to the fact that LDA potential decays exponentially rather than correctly as $-1/r$ for $r \rightarrow \infty$.
Therefore it is less binding for the outermost electrons. 

For the ground-states, it is seen that if the asymptotic behavior of the potential is improved,
$\epsilon_{max}$ becomes close to $E(N)-E(N-1)$. Two ways of making such a correction are the 
van Leeuwen and Baerends (LB)~\cite{leeuwen-baerends:1994} method and the range-separated hybrid (RSH) 
methods~\cite{inbook:savin:chong:1995,leininger-stoll-werner-savin:1997,iikura-tsundea-yanai-hirao:2001,yanai-tew-handy:2004,baer-neuhauser:2005,kronik-tamar-abramson-baer:2012}. 
In the LB method, a correction term is added to the LDA potential to make the effective potential go as $-1/r$ 
asymptotically, while in the RSH approach the Coulomb term is split into long-range (LR) and short-range (SR) part. 
Thus, $r^{-1}$ can be written as $ r^{-1} \operatorname{erf}(\gamma r) + r^{-1} \operatorname{erfc}(\gamma r )$ where $\gamma$ is a parameter~\cite{inbook:savin:chong:1995,leininger-stoll-werner-savin:1997,iikura-tsundea-yanai-hirao:2001,yanai-tew-handy:2004,baer-neuhauser:2005,kronik-tamar-abramson-baer:2012}. 
Here the first term is long-range and approaches $2\gamma/\sqrt{\pi}$ as $r\rightarrow 0$, while the second term is close to $\frac{\exp(-\gamma r)}{r})$~\cite{Bohm-Pines:1953c}  and is short range. 
In the RSH approach, the long-range part is treated exactly and the short-range part within the LDA. 
Recently, Stein et. al~\cite{stein-eisenberg-kronik-baer:2010}  have applied this idea to study the band gaps 
for a wide range of systems. In their work $\gamma$ is fixed by the satisfaction of the IP theorem. 
Motivated by their work, we have studied the IP theorem  using the LB potential. 
The line of our investigation is as follows: We first show that the LB potential leads to the satisfaction of the 
IP theorem for the ground-states to a high degree of accuracy. We then ask: does our approach of constructing excited-state 
energy functionals give the same level of accuracy for IP theorem for excited-states when applied with the LB potential ? 
This then provides  a test for our approach. 
The positive results of our calculations point to the correctness of our method of dealing with excited-state functionals.

The LB correction to the LDA potential is given as 
\begin{equation}
- \beta \rho_{\sigma}^{1/3}(\mathbf{r}) \frac{x^2_{\sigma}}{1+3\beta x_{\sigma} \sinh^{-1}(x_{\sigma})}
\label{eq:lbgradient}
\end{equation}
where the parameter $\beta$ is obtained by fitting the LB potential so that it resembles closely to the exact potential for the beryllium atom ($\beta=0.05$), and $x_{\sigma}$ is a dimensionless ratio given by 
$x_{\sigma} = \frac{|\nabla \rho_{\sigma}|}{\rho_{\sigma}^{4/3}}$.  
In the present paper, the parameter $\beta$ is  chosen  to satisfy IP theorem, similar to the work of Stein et al.~\cite{stein-eisenberg-kronik-baer:2010}.
The difference with the work of Ref.~\cite{stein-eisenberg-kronik-baer:2010} is that in the present work the potential 
is given entirely in terms of the density whereas in RSH functional it is written using both the wavefunction and the 
density. We note that recently the LB potential has also been applied to calculate satisfactorily the band gaps of a wide 
variety of bulk systems~\cite{Prashant-Harbola-etal:2013}.

In the following, we present in Section~\ref{sec:gr-theory} the results of application of the LB potential  
to the ground-states of several 
atoms. It is shown that with the help of parameter $\beta$, the LB potential can be optimized to satisfy IP theorem to a 
very high degree. 
The results for the ground-state set up the standard against which the excited-state results are to be judged for 
the functional and 
the corresponding potential proposed for the excited-states.
After this we study the IP theorem 
for excited-states using the LB correction in conjunction with the modified LDA potential based 
on the 
idea~\cite{inbook:Harbola-etal:Ghosh-Chattaraj:2013} of splitting $k$-space for excited-states. 
It is shown that the IP theorem is satisfied more accurately with the modified LDA potential 
in comparison to the ground-state LDA expression for the potential. In addition the modified potential has proper structure  
at the minimum of radial density in contrast to the ground-state LDA potential that has undesirable features at these 
points~\cite{cheng-Wu-Voorhis:2008}.

\section{Results for the ground-state IP theorem using LB potential}
\label{sec:gr-theory}
In this section, we first present the ground-state exchange-only $\epsilon_{max}$ and $\Delta$SCF energies obtained with the LDA and 
the LB potential 
for few atoms. Following that, we also present the results with correlation functional included.  The results for the ground-states are not
entirely new in light of some previous work~\cite{Banerjee-Harbola:1999} but it is necessary to give 
them here to put the new results of excited-states in proper perspective.  

The LDA exchange energy functional $E_{x}$~\cite{dirac30} is given by 
\begin{equation}
 E^{LDA}_{x}[\rho(\mathbf{r})]   
  = -\frac{3}{4} \left(\frac{3}{\pi} \right)^{1/3} \int \rho^{4/3} (\mathbf{r}) d\mathbf{r}  
  \label{eq:Ex-LDA}
\end{equation} 
and the corresponding potential $v^{LDA}_{x}$ required for self-consistency calculations is  
\begin{equation}
 v_{x}^{LDA} = - \left( \frac{6\rho(\mathbf{r})}{\pi} \right)^{1/3} 
 \label{eq:vx-LDA}
\end{equation}

Spin generalization of the expression of Eq.~\eqref{eq:Ex-LDA}, the local spin-density approximation (LSD), is obtained by using 
\begin{equation}
 E_x^{LSD} [\rho_{\alpha},\rho_{\beta}] = \frac{1}{2} E_x [2\rho_{\alpha}] + \frac{1}{2} E_x [2\rho_{\beta}].
\end{equation}
In Table~\ref{tab:gr-x-IP}, the $\epsilon_{max}$ and $\Delta$SCF obtained using the spin-generalized LDA exchange 
functional Eq.~\eqref{eq:Ex-LDA} and its potential Eq.~\eqref{eq:vx-LDA} is shown. 
As is well-known and noted earlier, the LSD underestimates the highest occupied orbital energy (HO) roughly by $50\%$, 
due to incorrect asymptotic exponential behavior of the LDA exchange-potential of Eq.~\eqref{eq:vx-LDA}.
The $\Delta$SCF energies, however, are close to the HF values. 
As stated in the previous section, $\epsilon_{max}$ and $\Delta$SCF energies become consistent with each other if  
asymptotically the potential goes correctly as $-1/r$.  
The van-Leeuwen and Barends (LB) potential does that.   

The LB potential $v_x^{LB}$~\cite{leeuwen-baerends:1994}, is calculated by including the LB correction of 
Eq.~\eqref{eq:lbgradient} to the LSD potential of Eq.~\eqref{eq:vx-LDA} and is given as 
\begin{equation}
v_{x,\sigma}^{LB}(\mathbf{r}) = v_{x,\sigma}^{LSD} - \beta \rho_{\sigma}^{1/3}(\mathbf{r}) \frac{x^2_{\sigma}}{1+3\beta x_{\sigma} \sinh^{-1}(x_{\sigma})}
\label{eq:vx-LB}
\end{equation}
where $v_{x,\sigma}^{LSD} = \frac{ \delta E_x^{LSD}}{\delta \rho_{\sigma}}$. 

In the original LB potential, parameter $\beta=0.05$. In the present work, in addition to using this value of $\beta$, 
we also optimize it by the satisfaction of IP theorem. 
In the latter calculation, $\beta$ is varied until $\epsilon_{max}$ and $\Delta$SCF energies match, i.e
\begin{equation}
 \epsilon^{\beta}_{max} = E(N,\beta) - E(N-1,\beta)
\end{equation}
Here, $\epsilon^{\beta}_{max}$ is the highest-occupied eigen-value for a specific choice of $\beta$. 
The price for employing the asymptotically corrected 
model exchange potential is that the corresponding   exchange functional is not known. 
Although in the past Levy-Perdew relation~\cite{levy-perdew:1985} has been used to get the corresponding exchange-energies from the 
potential~\cite{Banerjee-Harbola:1999}, this 
may not always be correct~\cite{Gaiduk-Chulkov-Staroverov:2009}. In this section we therefore use 
the potential above in the KS calculations but employ the 
LSD exchange  functional  for calculating the energies. 

Presented in Table~\ref{tab:gr-x-IP} are the results for $\epsilon_{max}$ and $\Delta$SCF energies using 
the LSD potential, the LB potential with $\beta=0.05$ and the LB potential with optimized $\beta$.  
As mentioned above, the exchange energy functional used is the LSD functional itself for all the three potentials. 
Also shown are the $\Delta$SCF energies obtained from HF calculations. 
Comparing the results of the LSD and LB calculations with the corresponding numbers in Hartree-Fock theory, it is evident 
that (i) the $\Delta$SCF values given by the LSD functional are reasonably close to the corresponding HF values, and 
(ii) by making the potential correct in the asymptotic regions, $\epsilon_{max}$ improves substantially and becomes 
close to the $\Delta$SCF values.  Interestingly, the match between $\epsilon_{max}$ obtained with the LB potential 
and the $\Delta$SCF values is better than that in the Hartree-Fock theory.

\begin{center}
\begin{figure}[h]
\includegraphics[width=4.5in,height=6.5in]{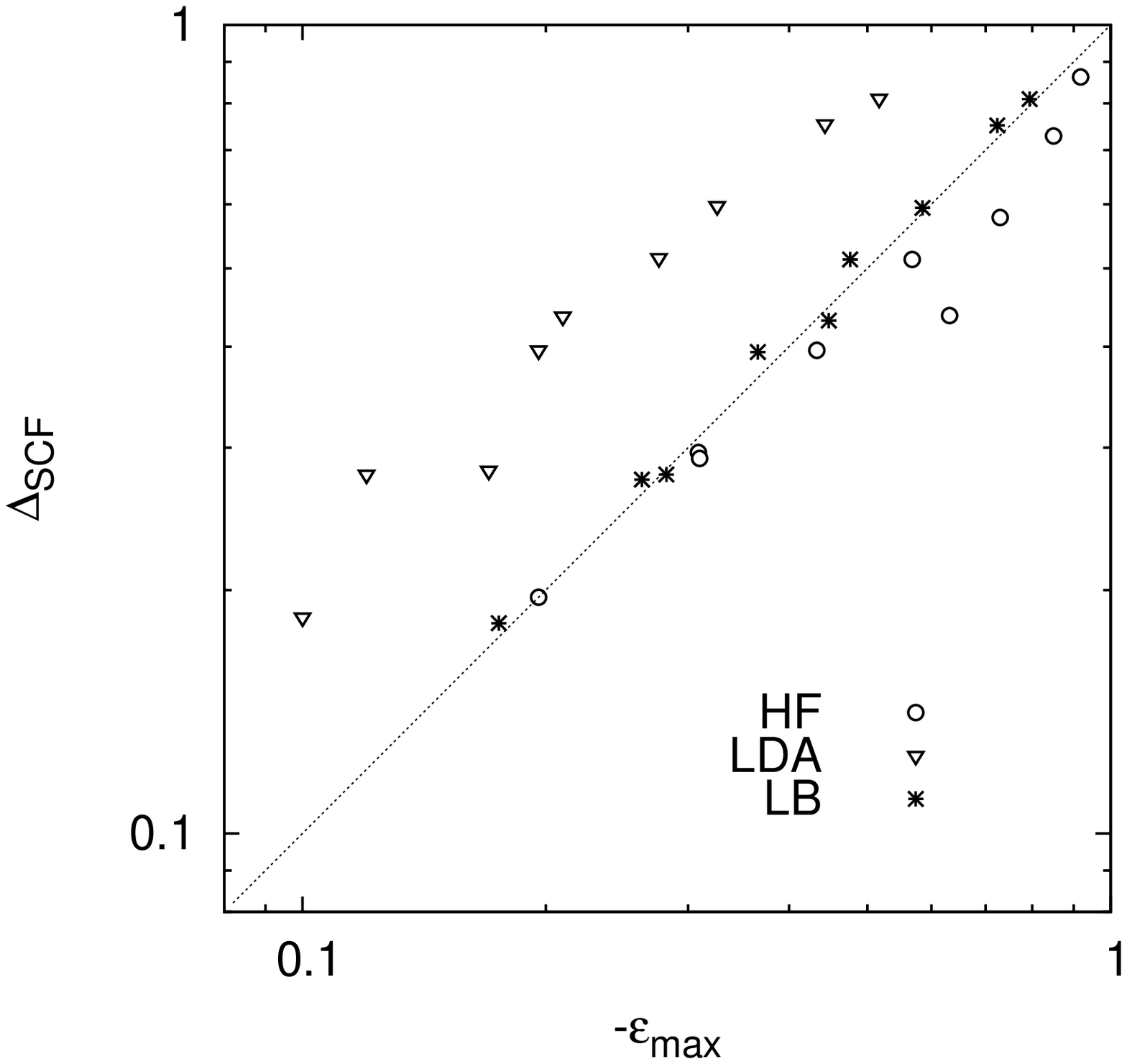}
\caption{Plot of $\epsilon_{max}$ vs. $\Delta$SCF energies using different exchange-only potentials.}
\label{fig:ip-gr-x}
\end{figure}
\end{center}

Next, motivated by the work of Ref.~\cite{stein-eisenberg-kronik-baer:2010}, we tune the parameter $\beta$ in the 
LB potential so that $\epsilon_{max}$ matches with the $\Delta$SCF energies. 
The optimized $\beta$ and the corresponding energies are also shown in Table~\ref{tab:gr-x-IP}. As is evident from 
Table~\ref{tab:gr-x-IP}, choosing   $\beta$ through IP theorem, the highest orbital energy $\epsilon_{max}$  
improves.  We note that according to Koopmans theorem~\cite{Koopmans:1934},
the orbital energy $\epsilon_{i}$ is close to the removal energy of the electron from that orbital. 
However we find that DFT results are better in this regard. 
The results of Table~\ref{tab:gr-x-IP} are depicted in Fig.~\ref{fig:ip-gr-x} where we have plotted the $\Delta$SCF results 
against $-\epsilon_{max}$ for LSD, LB and HF theories.  We see that the LB results are closest to the $\Delta$SCF$=-\epsilon_{max}$ line.

Having presented our results for the exchange-only calculations we next include  correlation using the LDA.
The correlation functional we use is that parametrized by Vosko, Wilk and Nusair~\cite{vosko-wilk-nusair:1980}. 
The orbital energies $\epsilon_{max}$ and the $\Delta$SCF energies for the  LB and $\beta$ optimized LB are presented in 
Table~\ref{tab:gr-xc-IP} in comparison with the experimental results~\cite{w2012crc}. 

We see from Table~\ref{tab:gr-xc-IP} that with the asymptotically corrected LB potential, the IP theorem is satisfied remarkably well.
The parameter $\beta$ in the LB potential is tuned to satisfy IP theorem (Ref. Eq.~\eqref{eq:ip})
The $\epsilon_{max}$, so obtained matches with experiments in a much better way.

\begin{center}
 \begin{figure}[h]
 \includegraphics[height=5.5in,width=3.8in]{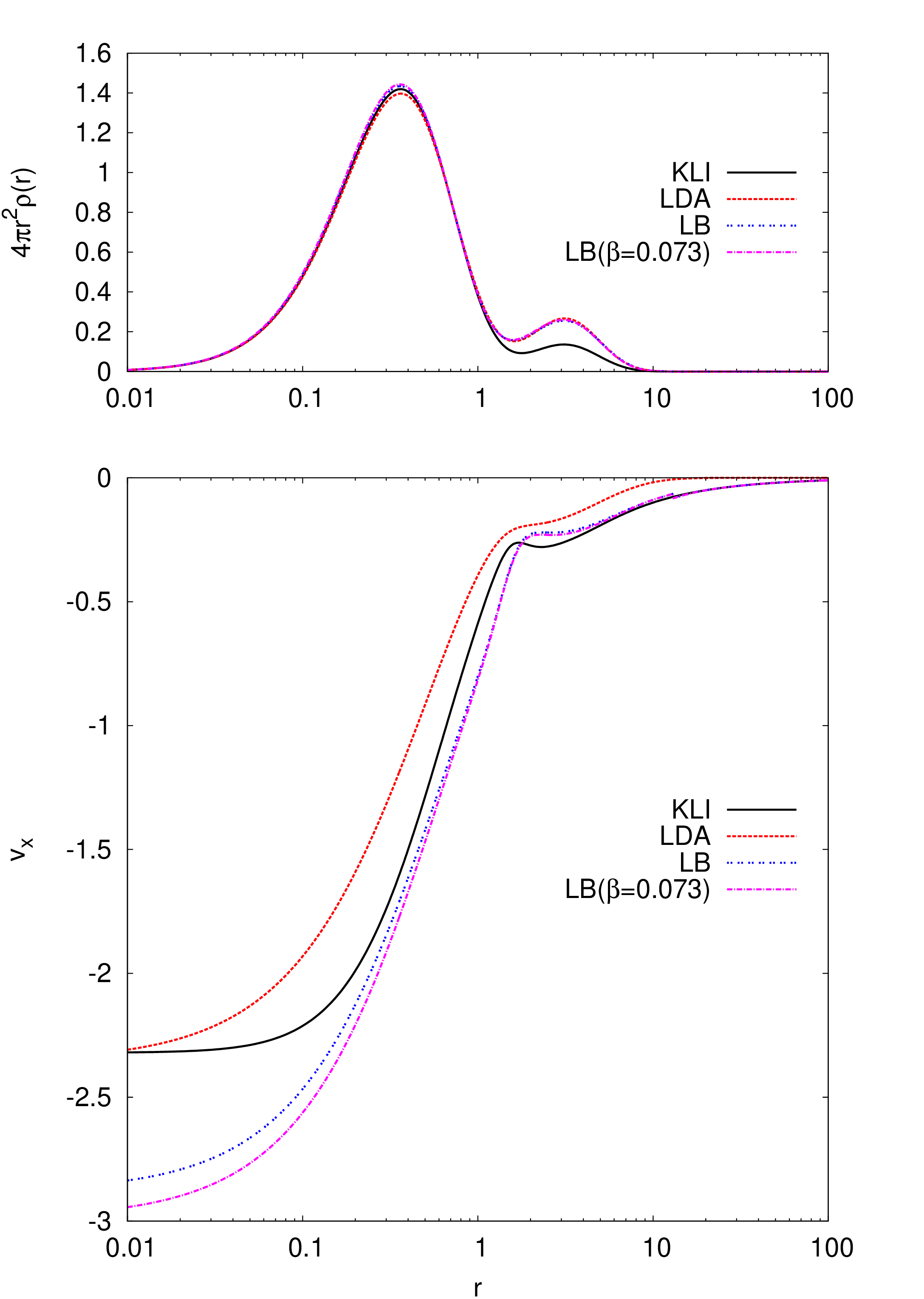}
 \caption{Ground-state radial density and the exchange potential of Li ($2s^1 \ 2S$) for the up spin obtained using 
 different approximations for the potential.}
 \label{fig:vx-Li-ground}
\end{figure}
\end{center}

The radial density  and the exchange potential for Li ground-state obtained using the LDA and the LB potentials 
are shown in Fig.~\ref{fig:vx-Li-ground}. Also shown in Fig.~\ref{fig:vx-Li-ground} is the KLI 
potential~\cite{krieger-li-iafrate:1992a}, which is essentially the exact exchange potential, for comparison. 
It is evident that from about $r=0.2 a.u.$ onwards, the LB potentials with both $\beta=0.05$ and the optimized $\beta$ are 
quite close to the KLI potential. The discrepancy of the LB potential for $r<0.2 a.u$ corresponds to the non-zero $\beta$. 
Furthermore, all the three potentials go as $-1/r$ in the asymptotic regions.  
On the other hand, the LSD potential underestimates the exact potential all over.
The bump in the potential for Li is at the minimum in the radial densities~\cite{lindgren:1971}.

Having given the results for the ground-states, we now turn our attention to excited-states and show that the exchange 
functional and potential constructed for these states by splitting the $k$-space for HEG  give results with similar accuracy.  

\section{Split $k$-space method for constructing excited-state energy functionals and excited-state potential}
\label{sec:excited}
\begin{center}
 \begin{figure}[h]
 \includegraphics[height=3.5in,width=4.5in]{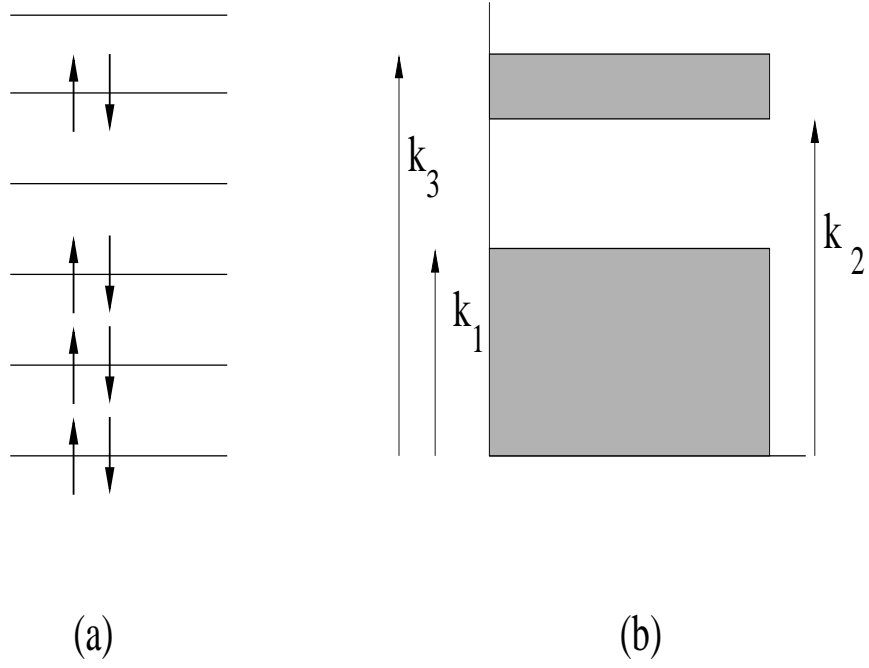}
 \caption{(a) Orbital and (b) the corresponding $k$-space occupation in the
excited state configuration of a homogeneous electron gas (HEG).}
 \label{fig:k-space}
\end{figure}
\end{center}

In eDFT, we have put forth the idea that the excited state energies be calculated using the modified local spin density (MLSD) 
functional developed over the past few years~\cite{samal-harbola:2005,shamim-harbola:2010,hemanadhan-harbola:2010,hemanadhan-harbola:2012}.
The basis of the MLSD exchange energy functional is the split $k$-space method~\cite{inbook:Harbola-etal:Ghosh-Chattaraj:2013}. 
In this method, 
the $k$-space is split in accordance to the orbital occupation of a given excited-state. 
In Fig.~\ref{fig:k-space}, we show an excited-state, where some orbitals (core) are occupied, followed by vacant (unocc) orbitals and 
then again orbitals are occupied (shell). 
To construct excited-state functionals, the density for each point is mapped onto the $k$-space of an HEG. 
The corresponding split $k$-space, also shown in Fig.~\ref{fig:k-space}, is constructed according to 
the orbital occupation i.e. the $k$-space is occupied from $0$ to $k_1$, vacant from $k_1$ to $k_2$ and then again occupied from $k_2$ to 
$k_3$ where $k_1, k_2, k_3$ are given by 
\begin{align}
 k_{1}^{3}(\mathbf{r}) 				&= 3\pi^{2}\rho_{c}(\mathbf{r})	\label{eq:k1}	\\
 k_{2}^{3}(\mathbf{r})-k_{1}^{3}(\mathbf{r}) 	&= 3\pi^{2}\rho_{v}(\mathbf{r})	\label{eq:k2}	\\
 k_{3}^{3}(\mathbf{r})-k_{2}^{3}(\mathbf{r}) 	&= 3\pi^{2}\rho_{s}(\mathbf{r})	\label{eq:k3}
\end{align}
in terms of  $\rho_{c}$, $\rho_{v}$ and $\rho_{s}$ corresponding to the electron densities of core, vacant (unoccupied) and the shell
orbitals.  Further,
\begin{align}
   \rho_{c}(\mathbf{r})	&= \sum\limits^{n_1}_{i=1} {\left| \phi_{i}^{core}(\mathbf{r})\right|}^{2}	\\
   \rho_{v}(\mathbf{r})	&=  \sum\limits^{n_2}_{i=n_1+1} {\left| \phi_{i}^{unocc}(\mathbf{r})\right|}^{2}	\\
   \rho_{s}(\mathbf{r})	&= \sum\limits^{n_3}_{i=n_2+1}  {\left| \phi_{i}^{shell}(\mathbf{r})\right|}^{2}
\end{align}
where first $n_1$ orbitals are occupied, $n_1+1$ to $n_2$ are vacant followed by occupied orbitals from $n_2+1$ to $n_3$.
The total electron density $\rho({\bf r})$ is given as
\begin{align}
 \rho({\mathbf{r}})=\rho_{c}({\mathbf{r}})+\rho_{s}({\mathbf{r}}) 	\\
 \textrm{or}  \hspace{2em}  \rho({\mathbf{r}})=\rho_1({\mathbf{r}}) -\rho_2({\mathbf{r}}) + \rho_3({\mathbf{r}})
\end {align}
with $\rho_1=\rho_c,\rho_2=\rho_c+\rho_v $ and  $\rho_3=\rho_c+\rho_v+\rho_s$. 
Using this idea, we have constructed the kinetic~\cite{hemanadhan-harbola:2010} and exchange-energy functionals~\cite{samal-harbola:2005} 
for excited-states, and shown that these functionals lead to accurate kinetic, exchange, and transition energies. 
We point out that the application of the ground-state LSD functional generally leads to poor results for 
excited-states. 
The generality of this idea to construct energy functionals for other class of systems also leads to 
accurate energies~\cite{shamim-harbola:2010,hemanadhan-harbola:2012}. 
Encouraged by these studies, we now subject this method to test by the IP theorem. 

 For this study, we consider the class of excited-systems as shown in Fig.~\ref{fig:k-space} for which 
the MLSD functional is given by~\cite{samal-harbola:2005}
\begin{align}
 E_X^{MLDA}[\rho] & =  \int \rho(\mathbf{r}) \left[ \epsilon(k_3)-\epsilon(k_2)+\epsilon(k_1) \right] d\mathbf{r}  
	      + \frac{1}{8\pi^3} \int \left(k_3^2-k_1^2 \right)^2 \ln \left( \frac{k_3+k_1}{k_3-k_1}\right) d\mathbf{r} \nonumber \\  
            &  - \frac{1}{8\pi^3} \int \left(k_3^2-k_2^2 \right)^2 \ln \left( \frac{k_3+k_2}{k_3-k_2}\right) d\mathbf{r}  
               + \frac{1}{8\pi^3} \int \left(k_2^2-k_1^2 \right)^2 \ln \left( \frac{k_2+k_1}{k_2-k_1}\right) d\mathbf{r} 
\label{eq:xmlda}
\end{align}
where $\epsilon(k_i)=\frac{-3k_i}{4\pi}$ is the exchange energy per particle for the ground state of HEG with Fermi wavevector $k_i$. 
Like the ground-state functional the modified local spin density (MLSD) functional is given as 
\begin{equation}
  E_X^{MLSD}[\rho] =  \frac{1}{2} E_X^{MLDA}[2 \rho_{\alpha}] + \frac{1}{2} E_X^{MLDA}[2 \rho_{\beta}] 
 \label{eq:exmlsd}
\end{equation}

The corresponding potential $v^{MLSD}_x$ is given as 
\begin{equation}
v_{x,\sigma}^{MLSD}(\textbf{r})= \frac{\delta E_{X}^{MLSD}[\rho]}{\delta\rho_{\sigma}(\mathbf{r})}
\label{xp}
\end{equation}
However, it has not been possible to get a workable analytical expression for $ v_{x}^{MLSD}({\bf r})$ out of 
Eqs.~\eqref{eq:xmlda} and~\eqref{xp}. 
Therefore on the basis of arguments based on ground-state theory, we try to model the potential. 
For completeness we note the earlier attempts to construct accurate excited-state potentials by Gaspar~\cite{gaspar:1974} 
and Nagy~\cite{nagy:1990}. They have given an ensemble averaged exchange potential for the excited states and 
using this potential, they calculate excitation energy for single electron excitations. 
In the next section we propose an excited-state LDA-like exchange  potential based on split $k$-space. 
This potential is similar to its ground-state LDA counterpart. We refer to this as the MLSD potential. 
We further correct the potential for its asymptotic behavior with the LB correction. With the asymptotically corrected MLSD potential, 
we show  that the IP theorem for excited-states is satisfied to a good accuracy.

\subsection{Generalization of Dirac exchange potential for excited-states using split $k$-space}
The Hartree-Fock exchange potential for a system of electrons is given by
\begin {equation}
v_{x,i}^{HF}=v_{x}(\phi_{i})=-\sum_{j}\int\frac{\phi^{\ast}_{j}({\bf r'})\phi_{i}({\bf r'})\phi_{j}({\bf r})}
{\phi_{i}({\bf r})|{\bf r}- {\bf r'}|}d{\bf r'}
\label {hfx}.
\end {equation}
For homogeneous electron gas, the wavefunction is given by
\begin {equation}
\phi_{{\bf k}}({\bf r}) = \frac{1}{{\sqrt V}} e^{ \left (i{\bf k}\cdot {\bf r}\right )} 
\label {wf}.
\end {equation}
where $V$ is the volume of the system. Using this form of wavefunction in Eq.~\eqref{hfx} we get an 
exchange potential for one-gap systems shown in Fig.~\ref{fig:k-space} to be for $\phi_k(\mathbf{r})$

\begin{align}
 v_{x}(k)=&-\frac{1}{\pi} \left [k_{1}-k_{2}+k_{3} + \frac {k_{1}^{2}-k^{2}}{2k}\ln\left|\frac{k+k_{1}}{k-k_{1}}\right| \right.  
  - \left. \frac {k_{2}^{2}-k^{2}}{2k}\ln\left|\frac{k+k_{2}}{k-k_{2}}\right|+\frac {k_{3}^{2}-k^{2}}{2k}\ln\left|\frac{k+k_{3}}{k-k_{3}}\right|\right ]
\label{eq:hfxpi}
\end{align}
where $k_1, k_2$, and $k_3$ are given by Eqns~\eqref{eq:k1},~\eqref{eq:k2},~\eqref{eq:k3}

This potential is orbital dependent. To make this potential an orbital independent potential we draw the analogy from the ground state exchange potential, where 
the exact LDA potential is equal to the HF potential for highest occupied molecular orbital (HOMO). 
\begin{equation}
 v^{HF}_{x,i}(\mathbf{r})|_{i=max} = \frac{\delta E_x^{LDA}[\rho(\mathbf{r})]}{\delta \rho(\mathbf{r})}
 \label{eq:hflda}
\end{equation}
Therefore we take the potential for the electron in HOMO as the exchange potential  
for all the electrons.  For this we put $k=k_{3}$ in Eq.~\eqref{eq:hfxpi},  
and get the following expression for the MLSD exchange potential 
\begin {align}
 v_{x}^{MLSD}= & -\frac{k_{3}}{\pi}\left [1-x_{2}+x_{1}-\frac{1}{2}(1-x_{1}^{2})\ln\left|\frac{1+x_{1}}{1-x_{1}}\right| \right.  
  \left. +\frac{1}{2}(1-x_{2}^{2})\ln\left|\frac{1+x_{2}}{1-x_{2}}\right|\right ]
\label {eq:vxmlsd}
\end {align}
where, $ x_{1}=\frac{k_{1}}{k_{3}}, x_{2}=\frac{k_{2}}{k_{3}} $

The MLSD potential of Eq.~\eqref{eq:vxmlsd} is also obtained by taking the functional derivative of 
the exchange functional $E_x^{MLDA}$ of Eq.~\eqref{eq:xmlda} with respect to $\rho_{3}(\mathbf{r})$,  
corresponding to the largest wave-vector in the $k$-space. 
Thus, we reach the same result from two different paths; this in some sense assures us about the correctness of the approach taken.  
When this potential is corrected for its asymptotic behavior by adding the LB correction, we obtain the modified LB (MLB) potential.

In the following Section, we test the MLB potential using the IP theorem for excited-states and show that it 
satisfies the IP theorem as accurately as the LB potential does for the ground-states.
On the other hand, the LB potential does not lead to as accurate as satisfaction of the IP theorem indicating thereby that the 
potential derived on the basis of splitting $k$-space is more appropriate for the excited-state calculations. 

\section{Results for excited-states}
\label{sec:ex-result}

The MLSD potential is the ground-state counterpart of the LSD potential. 
To correct the potential in the asymptotic region, 
we include the LB gradient term of Eq.~\eqref{eq:lbgradient} corresponding to largest wave-vector $k_3$ 
in the MLSD potential 
and obtain the MLB potential.
  \begin{equation}
   v_{x,\sigma}^{MLB} = v_x^{MLSD} - \beta \rho_{3,\sigma}^{1/3}(\mathbf{r}) \frac{x^2_{3,\sigma}}{1+3\beta x_{3,\sigma} \sinh^{-1}(x_{3,\sigma})} 
   \label{eq:vxmlb}
  \end{equation}
In performing self-consistent calculations, it is this potential that is employed as the exchange potential in the 
excited-state Kohn-Sham equations. 
Our calculations are performed using the central-field approximation~\cite{Slater:1929} whereby the potential is taken 
to be spherically symmetric. 
Having obtained the orbitals the exchange energy is then calculated 
using the MLSDSIC functional~\cite{samal-harbola:2005} 
and is 
given as 
\begin {equation}
E_{X}^{MLSDSIC}=E_{X}^{MLSD}-\sum_{i}^{rem}{E_i}^{SIC}-\sum_{i}^{add}{E_i}^{SIC}
\label{eq:mlsdsic}
\end {equation}
where,
\begin {equation}
E_{i}^{SIC}\left[\phi_i\right]= \int\int\frac{|\phi_{i}(\mathbf{r}_{1})|^{2}|\phi_{i}
({\bf r}_{2})|^{2}}{|{\bf r}_{1}-{\bf r}_{2}|}d{\bf r}_{1}d{\bf r}_{2} 
+E^{LSD}_{X}\left[\rho\left(\phi_i\right)\right]
\label{eq:sic}
\end {equation}
where the  summation index $i$ in Eq.~\eqref{eq:mlsdsic} runs over the orbitals from which the electrons are removed 
and create a gap, and to the orbitals to which the electrons are added. $E^{LSD}_{X}\left[\rho\left(\phi_i\right)\right]$ 
is the exchange energy corresponding to the $\phi_i$ orbital in the LSD approximation. 
Using the $\Delta$SCF energy obtained from these calculations and the eigenvalues from the Kohn-Sham calculations,
we study the IP theorem.
For our study, we have considered systems for which both the atomic excited-states and its ionic states can be represented by a single 
Slater determinant; this is so because LSD/MLSD is accurate for such states ~\cite{barth:1979}.

\begin{figure}
\centering
\includegraphics[width=4.5in,height=6.5in]{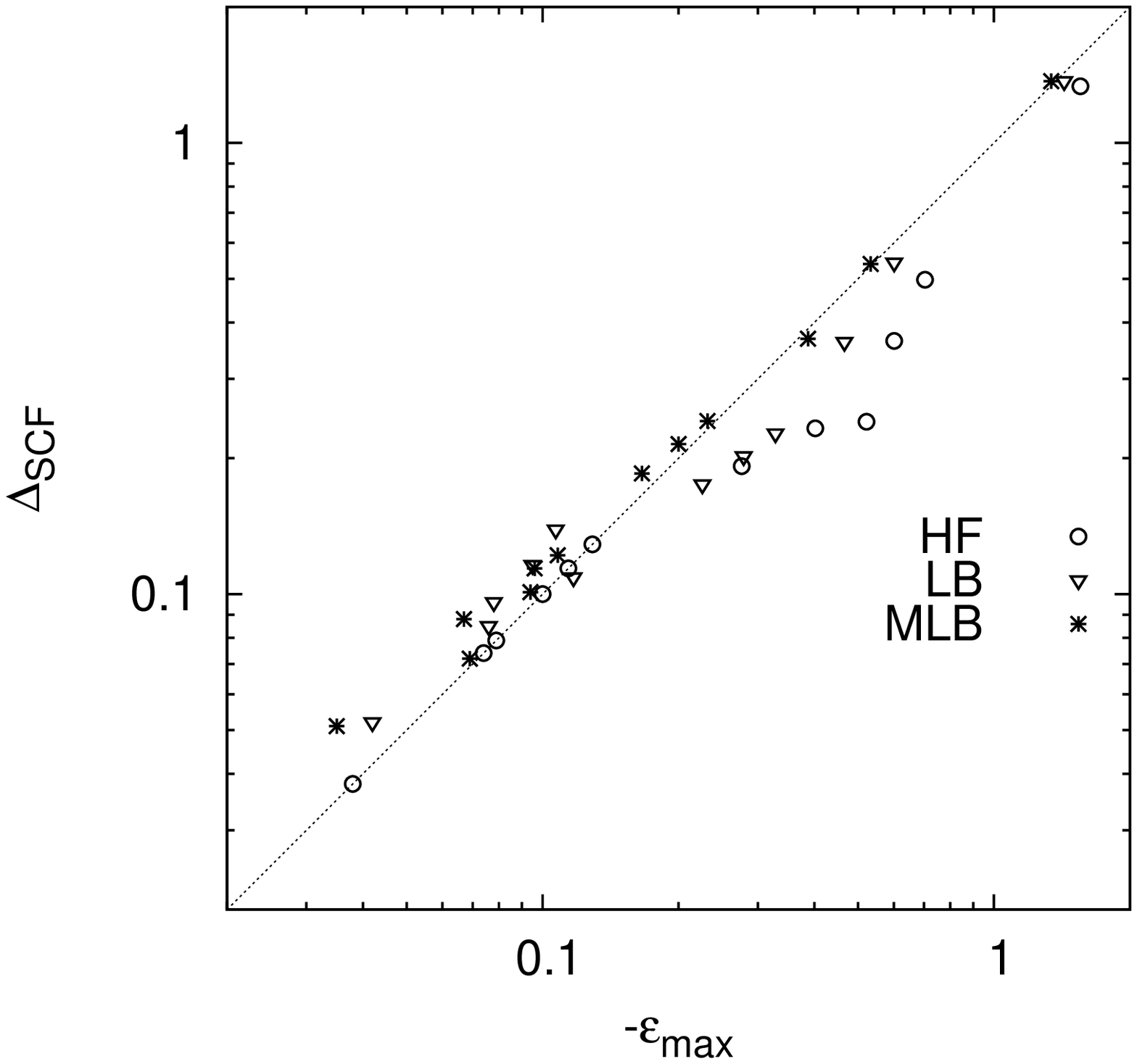}
\caption{Plot of $\epsilon_{max}$ vs. $\Delta$SCF energies using different exchange-only potentials of LB, MLB and HF.}
\label{fig:ip-ex-x}
\end{figure}

Presented in Table~\ref{tab:ex-x-IP} are the $\epsilon_{max}$ and $\Delta$SCF energies for different excited-states obtained using  
the LB potential of Eq.~\eqref{eq:vx-LB}, and the excited-state 
MLB potential of Eq.~\eqref{eq:vxmlb}. In both the LB and the MLB potentials, we have used $\beta = 0.05$.  Further, the energies
for both the potentials are calculated using the MLSDSIC exchange energy functional. 
The HF $\epsilon_{max}$ and $\Delta$SCF are also shown in Table~\ref{tab:ex-x-IP} for comparison. 
The results of Table III are shown graphically in Fig.~\ref{fig:ip-ex-x}.  
It is evident from the figure that the MLB potential satisfies the IP theorem accurately 
while the LB and HF both deviate from it. Thus accounting for the occupation of orbitals in the 
$k$-space gives better results for the theorem. Let us next check how does the MLB potential compare with 
the KLI potential for excited-states.

\begin{center}
 \begin{figure}[h]
 \includegraphics[height=5.5in,width=3.8in]{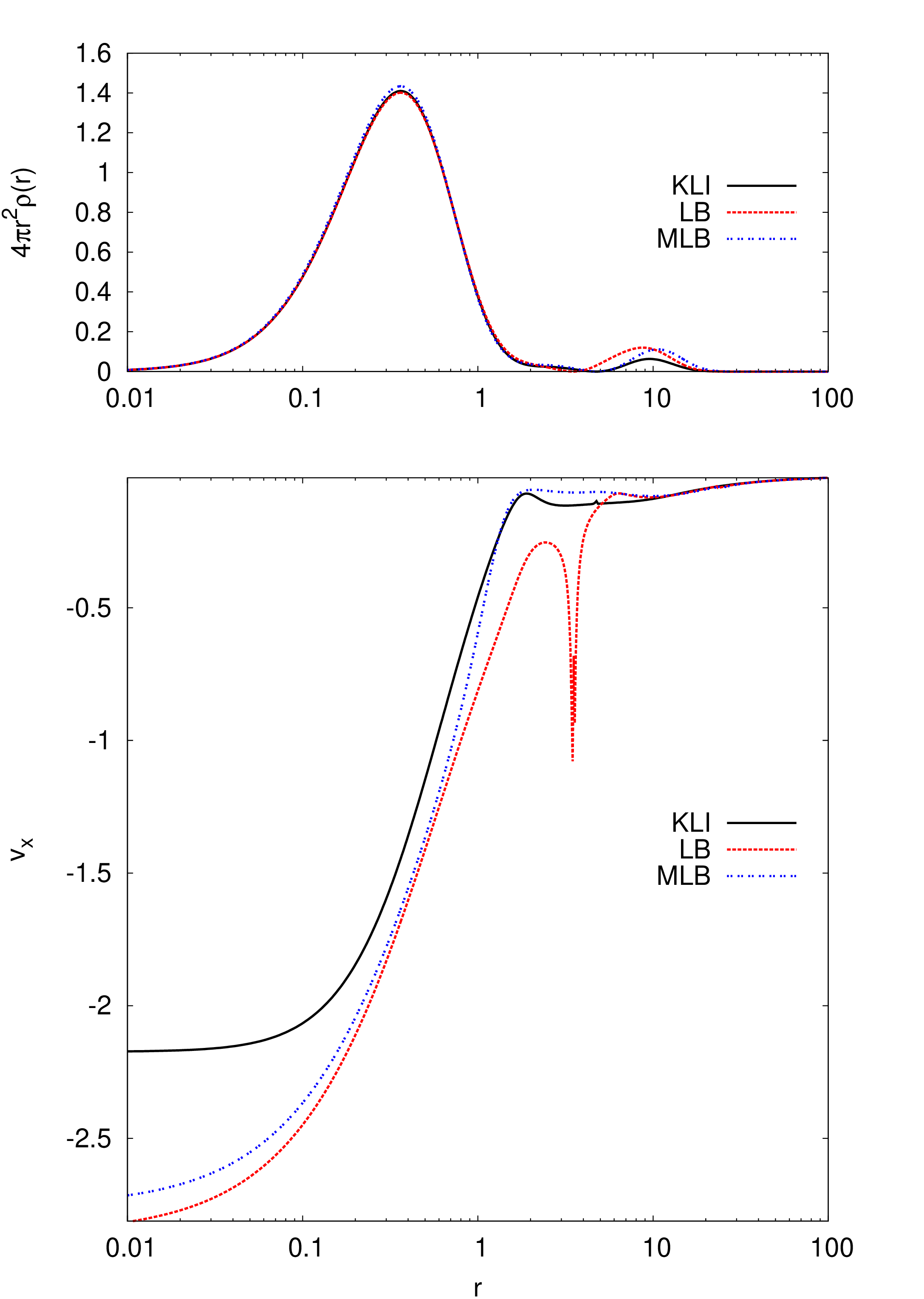}
 \caption{Excited-state radial density and the exchange potential of Li ($3s^1 \ 2S$) for the up 
 spin obtained using different approximations for the potential.}
  \label{fig:vx-Li-excited-3s1}
\end{figure}
\end{center}

Plotted in Fig.~\ref{fig:vx-Li-excited-3s1} are the radial density and the corresponding excited-state exchange 
potential of Li $(3s^1 \ ^2S)$ within the LB and the MLB approximations. 
Also shown in the figure the exact exchange potential, obtained through KLI method~\cite{Nagy:1997}.  
It is clear from the figure that the split $k$-space based MLB potential has a structure resembling the KLI potential for the excited-state: 
very close to it in the inter-shell region from about 0.1 a.u. onwards and beyond. 
This is similar to the relation between the LB potential and the KLI potential for the ground-states.
The LB potential for the excited-states, on the other hand, is
not close to the exact potential and has undesirable features at the minimum of radial density which are
not present in the MLB potential. 
Similar unsmooth behavior is observed~\cite{cheng-Wu-Voorhis:2008} in the LSD potential. 
In addition, the MLB potential is closer to the KLI potential in the interstitial and the asymptotic region, 
similar to what the LB potential did for the ground-states.
The discrepancy between the potentials near the nucleus that was present in the ground-states is also present here.  
Nonetheless it is clear that the exchange potentials obtained on the basis of split $k$-space give a much better 
description of an excited-state than the ground-state LB potential.

To sum up, we have shown that excited-state energy functional and its asymptotically corrected potential based on 
split $k$-space satisfy IP theorem with a great accuracy in the exchange-only limit. 
This can be improved further by optimizing $\beta$. 
In Table~\ref{tab:ex-x-IP}, we also present the results 
obtained by varying the parameter $\beta$ in the excited-state MLB potential until $\epsilon_{max}$ matches with the $\Delta$SCF energies. 
The $\epsilon_{max}$ so obtained using the excited-state potential is close to the HF values. 
For $B \ (3p^1 \ ^2P)$ we are unable to tune the $\beta$ using the MLB potential.

We now  wish to include correlation and compare our results with experiments.
The lack of correlation potential for excited-states forces us to rely on the ground-state potential. 
In Table~\ref{tab:ex-xc-IP} are the calculations performed using the ground-state VWN potential.
It is seen that similar to the ground-state, the $\Delta$SCF energies obtained with the split $k$-space functional 
are close to the experimental values. 
Also shown in table are the $\beta$ tuned energies to satisfy IP theorem. 
By imposing IP theorem, $\epsilon_{max}$ improves over the $\beta=0.05$ values and is closer to the 
experimental values for all atoms.
 
\section{Concluding Remarks}
\label{sec:conclusion}
To conclude we have shown that splitting $k$-space according to the occupation of Kohn-Sham orbitals is a 
good way of constructing excited-state potential. The potential so constructed, when corrected for its long-range behavior, 
gives highly accurate eigenvalues for the upper most orbital in the sense of IP theorem: the eigenvalues and the 
$\Delta$SCF energies obtained from the energy functional by splitting $k$-space agree with one another to a 
great degree. This shows that split $k$-space method could be the proper path to follow for constructing excited-state 
energy functionals.

\section{Acknowledgments}
M.~Hemanadhan wishes to thank Council of Scientific and Industrial Research (CSIR), New Delhi  for financial support.



\newpage 

\newpage
\newpage

\begin{table*}[Ht!]
\caption{Comparison of highest occupied eigen value $\epsilon_{\textrm{max}}$ and the $\Delta$SCF energies for atoms in their 
ground-state 
calculated using different exchange-only potentials. These are compared with the corresponding HF values. 
The numbers in $4^{th}$ and $5^{th}$ column are obtained with $\beta=0.05$ in Eq.~\eqref{eq:vx-LB} while those in column $7$ are obtained by 
optimizing $\beta$. The corresponding $\beta$ values are given in column $6$. 
 (All the energies are in a.u)}
\label{tab:gr-x-IP}
\centering
  \tabcolsep=0.11cm
\begin{tabular}{|l|c|c|c|c|c|c|c|c|}
\hline
Atoms/ion	 
&\multicolumn{2}{c|}{LDA}  
&\multicolumn{2}{c|}{LB($\beta=0.05$)} 
&\multicolumn{2}{c|}{LB($\beta$)}
&\multicolumn{2}{c|}{HF} \\
\cline{2-3}
\cline{4-5}
\cline{6-7}
\cline{8-9}
	 & $-\epsilon_{\textrm{max}}$ & $\Delta$SCF 
	 & $-\epsilon_{\textrm{max}}$ & $\Delta$SCF 
         & $\beta$	& $-\epsilon_{\textrm{max}} (=\Delta \textrm{SCF})$
	 & $-\epsilon_{\textrm{max}}$ & $\Delta$SCF 
          \\
\hline
He($1s^2 \ ^1S$)	&0.517	&0.811	&0.794	&0.810	&0.064	&0.809	&0.918	&0.862	\\
Li($2s^1 \ ^2S$)	&0.100	&0.185	&0.175	&0.182	&0.073	&0.182	&0.196	&0.196	\\
Be($2s^2 \ ^1S$)	&0.170	&0.281	&0.282	&0.278	&0.043	&0.278	&0.309	&0.296	\\
B($2s^2 2p^1 \ ^2P$)	&0.120	&0.278	&0.263	&0.274	&0.075	&0.273	&0.310	&0.291	\\
C($2s^2 2p^2 \ ^3P$)	&0.196	&0.396	&0.366	&0.394	&0.104	&0.392	&0.433	&0.396	\\
N($2s^2 2p^3 \ ^4S$)	&0.276	&0.515	&0.476	&0.513	&0.112	&0.511	&0.568	&0.513	\\
O($2s^2 2p^4 \ ^3P$)	&0.210	&0.436	&0.448	&0.431	&0.035	&0.432	&0.632	&0.437	\\
F($2s^2 2p^5 \ ^2P$)	&0.326	&0.597	&0.585	&0.594	&0.060	&0.594	&0.730	&0.578	\\
Ne($2s^2 2p^6 \ ^1S$)	&0.443	&0.754	&0.724	&0.751	&0.077	&0.749	&0.850	&0.729	\\
\hline
\end{tabular}
\end{table*}

\begin{table*}[h]
\caption{
Highest occupied eigen value $\epsilon_{\textrm{max}}$ and the $\Delta$SCF energies for atoms in  their ground-states
calculated using van Leeuwen and Baerends exchange and VWN correlation potential with the parameter $\beta=0.05$, and $\beta$ 
tuned, with the experimental values. (All the energies are in a.u)
}
\label{tab:gr-xc-IP}
\begin{tabular}{|l|c|c|c|c|c|}
\hline
Atom	 
&\multicolumn{2}{c|}{LB-VWN($\beta=0.05$)} 
&\multicolumn{2}{c|}{LB-VWN($\beta$)}  
 		 &Expt.~\cite{w2012crc} \\
\cline{2-3}
\cline{4-5}
	 & $-\epsilon_{\textrm{max}}$  & $\Delta$SCF 
         & $\beta$	& $-\epsilon_{\textrm{max}} (=\Delta \textrm{SCF})$
& \\
\hline
 He($1s^2 \ ^1S$)	&0.851	&0.892	&0.106	&0.890	&0.904	\\
 Li($2s^1 \ ^2S$)	&0.193	&0.198	&0.066	&0.198	&0.198	\\
 Be($2s^2 \ ^1S$)	&0.320	&0.329	&0.072	&0.329	&0.342	\\ 
 B($2s^2 2p^1 \ ^2P$)	&0.296	&0.312	&0.086	&0.311	&0.305	\\
 C($2s^2 2p^2 \ ^3P$)	&0.401	&0.431	&0.115	&0.430	&0.414	\\
 N($2s^2 2p^3 \ ^4S$)	&0.511	&0.550	&0.117	&0.548	&0.534	\\
 O($2s^2 2p^4 \ ^3P$)	&0.516	&0.506	&0.041	&0.507	&0.501	\\
 F($2s^2 2p^5 \ ^2P$)	&0.647	&0.661	&0.065	&0.660	&0.640	\\
 Ne($2s^2 2p^6 \ ^1S$)	&0.782	&0.813	&0.082	&0.811	&0.792	\\
\hline
\end{tabular}
 \end{table*} 

\small
\begin{table*}[h]
\caption{
Highest occupied eigen value $\epsilon_{\textrm{max}}$ and the $\Delta$SCF energies for atoms in different excited-states 
calculated using different exchange-only potentials. These are compared with the corresponding HF values. 
The numbers in $2^{nd}$ and $3^{rd}$ column are obtained with $\beta=0.05$ in Eq.~\eqref{eq:vx-LB}, in $4^{th}$ and 
$5^{th}$ column are obtained with $\beta=0.05$ in Eq.~\eqref{eq:vxmlb}, while in column $5$ are obtained by 
optimizing $\beta$. The corresponding $\beta$ values are given in column $4$. 
(All the energies are in a.u)
}
 \label{tab:ex-x-IP}
   \begin{tabular}{|l|c|c|c|c|c|c|c|c|c|c|}
\hline
Atom
& \multicolumn{2}{c|}{LB($\beta=0.05$)}
& \multicolumn{2}{c|}{MLB($\beta=0.05$)}
& \multicolumn{2}{c|}{MLB($\beta$)}
& \multicolumn{2}{c|}{HF}
\\
\cline{2-3}
\cline{4-5}
\cline{6-7}
\cline{8-9}
         & $-\epsilon_{\textrm{max}}$   & $\Delta$SCF
         & $-\epsilon_{\textrm{max}}$   & $\Delta$SCF
	 & $\beta$ & $-\epsilon_{\textrm{max}}(={\Delta \textrm{SCF})}$
         & $-\epsilon_{\textrm{max}}$   & $\Delta$SCF
         \\
\hline
Li($2p^1 \ ^2P$)        &0.117  &0.109  &0.096  &0.114	 &0.300	&0.114	&0.129  &0.129  \\
B($2s^1 2p^2 \ ^2D$)    &0.226  &0.175  &0.166  &0.185  &0.120	&0.183	&0.276  &0.192  \\
C($2s^1 2p^3 \ ^3D$)    &0.279  &0.202  &0.200  &0.215  &0.090	&0.213	&0.402  &0.233  \\
N($2s^1 2p^4 \ ^4P$)    &0.328  &0.227  &0.232  &0.242  &0.070	&0.241	&0.522  &0.241  \\
O($2s^1 2p^5 \ ^3P$)    &0.466  &0.362  &0.387  &0.368  &0.035	&0.370	&0.601  &0.364  \\
F($2s^1 2p^6 \ ^2S$)    &0.601  &0.543  &0.533  &0.539  &0.055	&0.538	&0.703  &0.497  \\
Ne$^+$($2s^12p^6\ ^2S$) &1.429  &1.369  &1.339  &1.370  &0.075	&1.369	&1.553  &1.334  \\
Li($3s^1 \ ^2S$)        &0.076  &0.085  &0.069  &0.072  &0.080	&0.073	&0.074  &0.074  \\
Li($4s^1 \ ^2S$)        &0.042  &0.052  &0.035  &0.051  &0.122	&0.038	&0.038  &0.038  \\
B($3s^1 \ ^2S$)         &0.107  &0.139  &0.108  &0.122  &0.200	&0.121	&0.114  &0.114  \\
B($3p^1 \ ^2P$)         &0.078  &0.096  &0.067  &0.088  &-	&-	&0.079  &0.079  \\
Be($2s^1 3s^1 \ ^3S$)   &0.095  &0.116  &0.094  &0.101  &0.100	&0.102	&0.100  &0.100  \\
\hline
\end{tabular}
 \end{table*}

\begin{table*}[h]
\caption{
Comparison of highest occupied eigen value $\epsilon_{\textrm{max}}$ and the $\Delta$SCF energies for atoms in various excited-states 
calculated using MLB exchange-only potential with VWN correlation potential with the experimental values. 
The numbers in $2^{nd}$ and $3^{rd}$ column are obtained with $\beta=0.05$ in Eq.~\eqref{eq:vxmlb} while those in column $5$ are obtained by 
optimizing $\beta$. The corresponding $\beta$ values are given in column $4$. 
(All the energies are in a.u)
}
 \label{tab:ex-xc-IP}
\begin{tabular}{|l|c|c|c|c|c|}
\hline
Atom
    &\multicolumn{2}{c|}{MLB-VWN($\beta=0.05$)}
    &\multicolumn{2}{c|}{MLB-VWN($\beta$)}
    &
\\
\cline{2-3}
\cline{4-5}
         & $-\epsilon_{\textrm{max}}$   & $\Delta$SCF
         & $\beta$	& $-\epsilon_{\textrm{max}} (=\Delta \textrm{SCF})$
         &Expt.~\cite{NIST} \\
\hline
 Li($2p^1 \ ^2P$)               &0.110  &0.128  &0.230  &0.127  &0.130  \\
 B($2s^1 2p^2 \ ^2D$)           &0.214  &0.252  &0.600  &0.247  &0.257  \\
 C($2s^1 2p^3 \ ^3D$)           &0.262  &0.300  &0.500  &0.295  &0.318  \\
 N($2s^1 2p^4 \ ^4P$)           &0.308  &0.344  &0.350  &0.339  &0.348  \\
 O($2s^1 2p^5 \ ^3P$)           &0.453  &0.441  &0.040  &0.442  &0.471  \\
 F($2s^1 2p^6 \ ^2S$)           &0.594  &0.604  &0.056  &0.600  &0.623  \\
 Ne$^+$($2s^1 2p^6 \ ^2S$)      &1.409  &1.444  &0.080  &1.443  &1.442  \\
 Li($3s^1 \ ^2S$)               &0.079  &0.081  &0.060  &0.081  &0.074  \\
 Li($4s^1 \ ^2S$)               &0.042  &0.046  &0.122	&0.046      &0.039  \\
 B($3s^1 \ ^2S$)                &0.123  &0.136  &0.200	&0.136      &0.122  \\
 B($3p^1 \ ^2P$)                &0.079  &0.099  &-	&-      &0.083  \\
 Be($2s^1 3s^1 \ ^3S$)          &0.107  &0.112  &0.082  &0.112  &0.105  \\
\hline
\end{tabular}
 \end{table*}

\end{document}